\documentclass[preprint]{vgtc}               

\graphicspath{{figures/}{pictures/}{images/}{./}} 

\usepackage{times}                     

\usepackage{tabu}   
\usepackage{amsmath}
\usepackage{booktabs}                  
\usepackage{lipsum}                    
\usepackage{mwe}                       
\usepackage{multirow} 
\usepackage{mathrsfs}
\usepackage{color}
\usepackage{amsmath}
\usepackage{amsfonts}
\usepackage{graphicx}  
\usepackage{amssymb}
\usepackage{multirow}
\usepackage[ruled,vlined]{algorithm2e}
\usepackage{mathrsfs}

\usepackage{mathptmx}                  

\newcommand\blfootnote[1]{%
  \begingroup
  \renewcommand\thefootnote{}\footnote{#1}%
  \addtocounter{footnote}{-1}%
  \endgroup
}

\onlineid{1010}

\vgtccategory{Research}

\vgtcinsertpkg

\preprinttext{To appear in  IEEE Workshop on Uncertainty Visualization in conjunction with IEEE VIS 2024, Florida, USA.}

\title{Uncertainty-Informed Volume Visualization using Implicit Neural Representation}

\author{Shanu Saklani\thanks{e-mail: shanu@cse.iitk.ac.in}\\ %
        \scriptsize IIT Kanpur, India %
\and Chitwan Goel\\ %
     \scriptsize IIT Kanpur, India %
\and Shrey Bansal\\ %
     \scriptsize IIT Kanpur, India %
\and Zhe Wang\\ %
     \scriptsize Oak Ridge National Lab., USA %
\and Soumya Dutta\thanks{e-mail: soumyad@cse.iitk.ac.in (Corresponding author.)}\\ %
     \scriptsize IIT Kanpur, India %
\and Tushar M. Athawale\\ %
     \scriptsize Oak Ridge National Lab., USA %
\and David Pugmire\\ %
     \scriptsize Oak Ridge National Lab., USA %
\and Christopher R. Johnson\\ %
     \scriptsize University of Utah, USA}

\DeclareUnicodeCharacter{2212}{\ensuremath{-}}

\newcommand{\bmark}[1]{#1}

\newenvironment{tight_enumerate}{
\begin{enumerate}
  \setlength{\itemsep}{0pt}
  \setlength{\parskip}{4.5pt}
}{\end{enumerate}}

\abstract{The increasing adoption of Deep Neural Networks (DNNs) has led to their application in many challenging scientific visualization tasks. While advanced DNNs offer impressive generalization capabilities, understanding factors such as model prediction quality, robustness, and uncertainty is crucial. These insights can enable domain scientists to make informed decisions about their data. However, DNNs inherently lack ability to estimate prediction uncertainty, necessitating new research to construct robust uncertainty-aware visualization techniques tailored for various visualization tasks. In this work, we propose uncertainty-aware implicit neural representations to model scalar field data sets effectively and comprehensively study the efficacy and benefits of estimated uncertainty information for volume visualization tasks. We evaluate the effectiveness of two principled deep uncertainty estimation techniques: (1) Deep Ensemble and (2) Monte Carlo Dropout (MCDropout). These techniques enable uncertainty-informed volume visualization in scalar field data sets. Our extensive exploration across multiple data sets demonstrates that uncertainty-aware models produce informative volume visualization results. Moreover, integrating prediction uncertainty enhances the trustworthiness of our DNN model, making it suitable for robustly analyzing and visualizing real-world scientific volumetric data sets.
} 


\begin{document}


\firstsection{Introduction}
\maketitle
The scientific visualization community has witnessed myriad applications of deep learning owing to the rapid growth of deep neural network (DNN) research~\cite{dl4scivis}. Among many applications, analysis, visualization, and representation of volumetric data using DNNs have emerged as a promising research domain. DNNs have been used effectively to (1) synthesize volume rendered images given rendering and view parameters~\cite{GANvolren}, (2) generate high-quality super-resolution images~\cite{medvol_superreso}, (3) perform interactive volume visualization~\cite{DNNVolVis}, (4) build differentiable rendering models to perform automatic viewpoint and transfer function optimization~\cite{weissDiffDVR},  (5) learn adaptive volume sampling strategies to produce accurate images~\cite{WeissSamplingVolren}, and (6) learn compressed volume representations~\cite{ levine_neural_compression, fvsrn}. While the above DNN-based approaches facilitate  multifaceted exploration of volumetric data, none of the approaches study the impact of prediction uncertainty associated with such models. Model prediction uncertainty, if reliably estimated and communicated to the domain experts, can enable them to make informed decisions about the data during analysis phase~\cite{Bonneau2014,gagh16}. \blfootnote{This manuscript has been authored by UT-Battelle, LLC under Contract No. DE-AC05-00OR22725 with the U.S. Department of Energy. The publisher, by accepting the article for publication, acknowledges that the U.S. Government retains a non-exclusive, paid up, irrevocable, world-wide license to publish or reproduce the published form of the manuscript, or allow others to do so, for U.S. Government purposes. The DOE will provide public access to these results in accordance with the DOE Public Access Plan (\url{http://energy.gov/downloads/doe-public-access-plan}).} As a consequence, the DNN-based volume visualization and analysis methods will become more trustworthy and reliable. However, a literature survey reveals that \bmark{epistemic uncertainty-aware DNN-based volume visualization needs detailed exploration - a gap that this work attempts to fill.} 

Given the recent success of implicit neural representations (INRs) for compact modeling and representing volumetric data sets~\cite{coordnet, levine_neural_compression, stsrinr}, in this work, we study the efficacy and applicability of uncertainty-aware INRs for volume visualization task. Since traditional DNNs do not quantify their prediction uncertainty, we augment our INRs with two different deep uncertainty quantification techniques to collect uncertainty estimates along with the predicted data values. We thoroughly analyze and visualize the estimated uncertainty to comprehend model accuracy, trustworthiness, and reliability.

Given the existing deep uncertainty estimation techniques, in this work, we prefer the methods of uncertainty estimation that can be incorporated into an existing DNN with minimal architecture modification so that the visualization community can readily adopt such models to conduct uncertainty-aware volume visualization using DNNs. To that end, we employ Deep Ensembles as our first uncertainty estimation method since Deep Ensembles often outperform other uncertainty estimation methods and produce more accurate predictions~\cite{ beluch2018power, gustafsson2020evaluating, huzw23}. However, the benefits of Deep Ensembles come at the cost of significantly large training time and resource requirements since multiple DNN models need to be trained to produce an ensemble of DNNs. To mitigate this challenge, we study the accuracy and viability of a single-model-based deep uncertainty estimation technique known as Monte Carlo Dropout (MCDropout)~\cite{gagh16}. Theoretically, the MCDropout is equivalent to approximate inferencing in deep Gaussian processes~\cite{dala13,gagh16}, making it an attractive choice for our work. 


We analyze the accuracy, effectiveness, and benefits of uncertainty-aware implicit neural representations (INRs) of volume data augmented with two deep uncertainty estimation techniques: (1) Deep Ensemble and (2) MCDropout method to carry out informative visual analysis of scientific data. \bmark{Since our methods provide model uncertainty for predicted data values, we are able to quantify fine-grained pixel-wise uncertainty for the volume-rendered images for any user-provided transfer function. As the user changes the transfer function, our pixel-wise uncertainty map also gets updated according to the current transfer function configuration.} We can further investigate how uncertainty impacts each color channel while computing the final pixel color using the ray casting method. 
Finally, we conduct a comprehensive study of the two principled deep uncertainty estimation methods to evaluate their applicability and viability in scientific volume visualization tasks. \bmark{We examine whether the error and uncertainty maps show any correlation and further advocate that when error computation is not possible due to the unavailability of ground truth data, model uncertainty is still available and can be helpful in assessing the trustworthiness of the volume rendering results.} Therefore, our contributions are twofold:
\begin{tight_enumerate}
	\item We propose the use of uncertainty-aware implicit neural representations for informative and robust visual analysis of volume data, emphasizing the significance of visualizing prediction uncertainty to aid domain scientists in interpreting their results more reliably.
	\item We thoroughly compare and contrast between two deep uncertainty estimation techniques: (1) Deep Ensemble, and (2) MCDropout, showcasing their applicability for conducting uncertainty-informed volume visualization.
\end{tight_enumerate}

\section{Background and Uncertainty in Deep Neural Networks}
\label{relworks}
First we summarize research works related to deep learning for scientific visualization and uncertainty visualization in the following section. Then, we briefly introduce different methods for quantifying uncertainty in deep neural networks, followed by a detailed description of two well-known methods of estimating uncertainty in DNNs used in this work.

\subsection{Deep Learning for Scientific Visualization}
Deep learning has found numerous applications in scientific visualization. Techniques for generating compact neural representations of scientific data are proposed by Lu et al.~\cite{levine_neural_compression} and Weiss et al.~\cite{fvsrn}. Hong et al.~\cite{DNNVolVis}, He et al.~\cite{insitunet}, and Berger et al.~\cite{GANvolren} study the visualization of scalar field data using volume-rendered images, and Weiss et al.~\cite{weiss2019volumetric} study volume data with isosurfaces. An adaptive sampling-guided approach is further used by Weiss et al.~\cite{WeissSamplingVolren} for efficient volume data visualization using DNNs. Another prominent area of research involves generation of spatiotemporal super-resolution volumes from low-resolution data using deep learning methods~\cite{han2020ssr, han2019tsr, wurster2022deep, han2021stnet}. For compressing the volume data, new models are proposed for domain-knowledge-aware latent space generation for scalar data~\cite{shen2022idlat}. For the generation of visualization and exploration of parameter spaces for ensemble data, DNNs are also used as surrogates~\cite{insitunet, shi2022gnn, shi2022vdl}. Han et al.~\cite{v2v} propose a variable-to-variable translation technique for scientific data. A comprehensive review of deep learning applications for scientific visualization are available in the state-of-the-art survey~\cite{dl4scivis}.

\subsection{Uncertainty Visualization}
Visualization of uncertainty continues to be an important and challenging area. Pang et al.~\cite{pang1997approaches} give one of the earlier summaries of uncertainty visualization techniques.
Visualization methods that are enhanced with tools for uncertainty estimation are given by Brodlie et al.~\cite{dill2012expanding}. Non-parametric models are used by Athawale et al.~\cite{athawale2020direct} to improve uncertainty in volume rendering. Spatial probability distributions, defined over triangular meshes, are visualized by Potter et al.~\cite{potter2008}, preceded by a classification of uncertainty visualization techniques~\cite{dienstfrey2012uncertainty}.

Uncertainty visualization techniques are also explored for isocontouring methods. The level crossing probability of adjacent points is computed by Pöthkow et al.~\cite{51187755}, and to calculate the level crossing probability for each cell, this method is further refined in~\cite{Pothkow2011}. In~\cite{athawale2022fiber}, a visual analysis of fiber uncertainty is done. Uncertainty visualization techniques are summed up in a comprehensive survey by Bonneau et al. in~\cite{Bonneau2014}. For image processing applications~\cite{gillmann2018uncertainty} and medical imaging~\cite{gillmann2021uncertainty}, uncertainty visualization techniques are summarized by Gillmann and colleagues. Whitaker et al.~\cite{whitaker2013contour} use contour boxplots to examine uncertainty visualization in an ensemble of contours. The latest progress regarding uncertainty in visualization research can be found in~\cite{kamal}.

\subsection{Uncertainty in Deep Neural Networks}
Despite the widespread popularity of deep learning techniques, there are significant concerns regarding their interpretability, robustness, and generalizability in real-world applications~\cite{zhang2021understanding}. The inability of conventional DNNs to provide uncertainty estimates can undermine the practical results they offer in fields such as computer vision, natural language processing, scientific visualization, and visual analytics. Developing universal techniques to measure and quantify uncertainty in DNNs remains a challenge, as the types and sources of uncertainty vary greatly across different applications, making it difficult to generalize uncertainty estimation methods. In the following, we succinctly describe some causes of uncertainty in DNNs and various methods proposed to simulate and mitigate them.

Uncertainty in a DNN can be broadly categorized into data (aleatoric) uncertainty and model (epistemic) uncertainty \cite{blundell2015weight, deepEnsembles}. Data uncertainty arises from errors and noise in measurement systems. In contrast, several factors contribute to model (epistemic) uncertainty. Firstly, current DNN models simplify real-world systems to generate observations, but this simplification introduces errors and uncertainty in predictions. Secondly, many DNNs require careful adjustments, such as implementing dropout~\cite{gagh16, hinton2012improving}, experimentally adjusting the model hyperparameters~\cite{you2019large}, and applying regularization~\cite{choi2019empirical}. Variations in these parameter settings can result in different outcomes and associated uncertainty.

\subsection{Methods to Model Uncertainty in DNNs}
In the following, we discuss various techniques used for modeling uncertainty in DNNs. Then, we discuss two well-known uncertainty estimation methods, MCDropout and Deep Ensembles in detail since these two methods are employed in our work.

\textbf{Deterministic Methods.}
Uncertainty estimation capabilities can be integrated into deterministic models by training a network explicitly to quantify uncertainties, as discussed in the work on evidential neural networks~\cite{evidential.neural.networks}.

\textbf{Test-time Augmentation Methods.}
Data augmentation during testing improves model performance using adversarial examples~\cite{Ayhan.2018, wang2019aleatoric} which determine prediction uncertainty.

\textbf{Deep Evidential Regression.} In deep evidential regression, the network simultaneously learns both parameters and hyperparameters for the corresponding evidential distributions~\cite{deep_evidential_uncert}. These evidential distributions are then utilized to estimate model uncertainty.

\textbf{Uncertainty via Stochastic Data Centering.} In this approach, the authors propose that training an ensemble of DNNs with data sets shifted by a constant bias allows for estimating model uncertainty by assessing the variability across predictions from these ensemble members. They further introduce a method to achieve similar uncertainty estimation using a single DNN~\cite{single_model_uncert}.

\textbf{Bayesian Methods.} Bayesian neural networks (BNNs) utilize prior distributions on the model parameters of DNNs to quantify epistemic uncertainty~\cite{BNNSurvey, deepEnsembles}. Training these networks typically are computatinally expensive and involves methods such as stochastic gradient MCMC~\cite{mafb17} and variational inference~\cite{hahn19}.

\begin{figure}[tb]
\centering
\includegraphics[width=\linewidth]{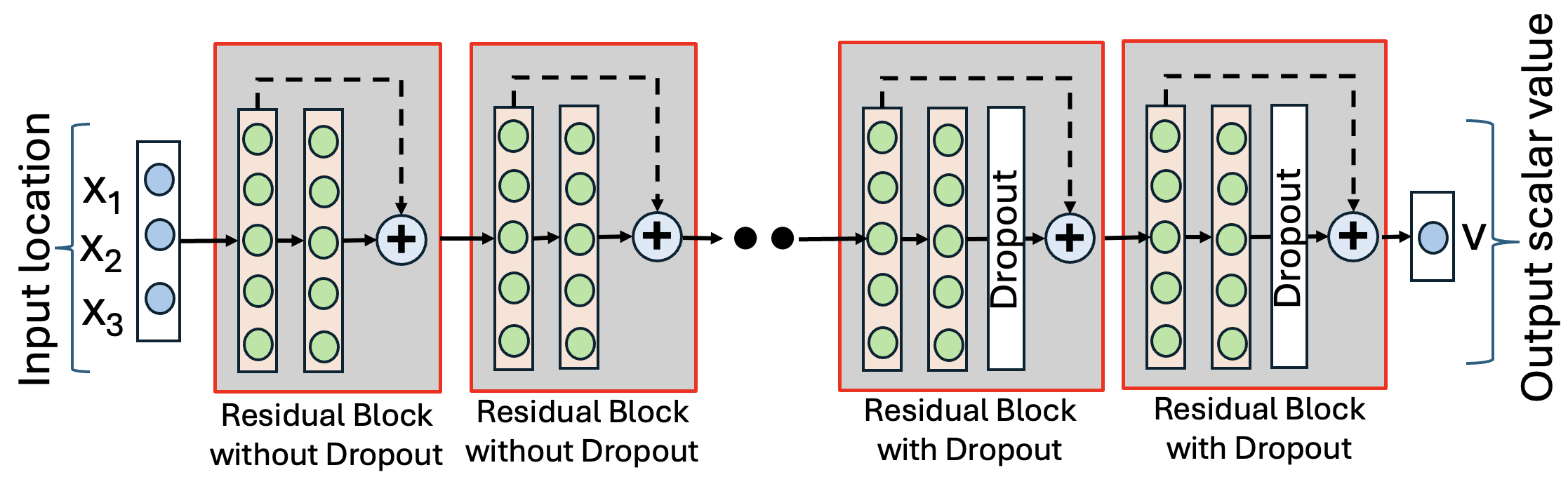}
\caption{The schematic of the MCDropout-enabled INR model which uses a residual block-based MLP architecture. Dropout layer is added at the last two residual blocks to generate uncertainty estimates during inference time. The INR architecture for the Ensemble method is identical to this, except there are no dropout layers.}
\label{model_arch}
\end{figure}

\subsection{Ensemble Method}
Ensemble techniques work on the idea that a group of learners of comparable competence is commonly better than a single learner~\cite{sagi2018ensemble}. Ensemble methods provide an inherent way to measure predictive uncertainty by assessing the differences among various predictions~\cite{leutbecher2008ensemble, parker2013ensemble}, besides just enhancing the generalization error. Therefore, the predictive uncertainty of DNNs can be calculated by taking ensemble methods into account~\cite{gustafsson2020evaluating,  renda2019comparing}. Deep Ensembles~\cite{deepEnsembles} are proposed by Lakshminarayanan et al., where DNNs are designed with two heads that give both the prediction and the associated uncertainty. In essence, Deep Ensemble learning can be equivalent to an approximation of Bayesian averaging~\cite{UQEnsemble}. In Bayesian averaging, the prediction of the final model is given as follows:
\begin{equation}
prediction = \int P_W(x)~\pi(W | D) \label{bayes_eqn}
\end{equation}
Here, \( P_W(x) \) denotes the probabilistic prediction for sample \( x \), and \( \pi(W|D) \) is the posterior probability distribution on the neural network weights, and \( D \) is the training data. In practice, estimation the above integral is very challenging, and exploration of all the modes of \( \pi(W|D) \) is not needed to calculate this integral precisely~\cite{UQEnsemble}. This observation suggests that, from a set of ensemble members, averaging predictions without weights can be viewed as an approximation of Equation~\ref{bayes_eqn}~\cite{UQEnsemble}. It should be noted that randomizing the training data and initializing parameters randomly introduces enough variety in each learned ensemble member to effectively predict the uncertainty during the training process, a technique adopted in this work to generate the Deep Ensemble model. It is worth mentioning that Ensemble techniques often surpass the techniques that depend on probabilistic back-propagation and Monte Carlo dropout~\cite{gustafsson2020evaluating, beluch2018power, ovadia2019can, vyas2018out}, and hence Ensemble methods are often regarded as the state-of-the-art. They also show more resilience to changes in data distribution.

\subsection{MCDropout Method}
Dropout~\cite{gagh16, hinton2012improving} is primarily used as a regularization technique, which is applied for fine-tuning machine learning models and preventing overfitting by optimizing the adjusted loss function. Gal et al., in their seminal work~\cite{gagh16}, propose a new viewpoint on how dropout can be efficiently used as a method for approximate Bayesian inference in deep Gaussian processes. By using dropout at test time and running many forward passes with different dropout masks, the model generates a range of predictions rather than a single point estimate and is equivalent to sampling from the Bayesian posterior distribution, $P(W \mid X, Y)$, where $W$ is the weights of the neural network model, $X$ is training data, and $Y$ is target output. The mean of these sampled predictions acts as the expected output of the model. Then, the epistemic uncertainty of the DNN can be conveniently estimated by calculating the standard deviation among these sampled predictions~\cite{shi2022gnn}. By collecting Monte Carlo (MC) samples from the network, which is dropout-enabled, such probabilistic predictions are obtained by running multiple forward passes at the time of inference, which is known as the Monte Carlo Dropout (MCDropout) method.


\section{Uncertainty-Aware Implicit Neural Representation of Scalar Field Data}
\label{method}
\subsection{Implicit Neural Representation}
Implicit neural representations (INRs) using periodic activation functions have shown promising results for learning representations of coordinate-based data sets. Input coordinates within the data domain are mapped to their associated output values using such neural networks. By using a sinusoidal activation function in a feed-forward neural network, known as SIREN (sinusoidal representation network), INRs can be effectively constructed, as proposed by Sitzmann et al.~\cite{siren} in their research. Recent works have employed many variations of SIREN to address multiple complex problems in the scientific data visualization community, achieving state-of-the-art results~\cite{levine_neural_compression, coordnet, stsrinr, neuralflowmap}. The achievements of these recent research efforts have inspired us to build our uncertainty-aware model by applying SIRENs as the base neural network architecture.

\subsection{Model Architecture}
Our objective is to learn a function that represents the mapping from the input data coordinate domain to the corresponding scalar value domain using an implicit neural network (INN). We make our base model a multilayer perceptron for this purpose. It has \( d \) input neurons (\( d \) can be 2 or 3 depending on the dimensionality of the scalar field being modeled), \( l \) hidden layers, and $1$ neuron in the output layer. As proposed in~\cite{resnet}, we strengthen the foundational SIREN architecture by combining residual blocks and skip connections to enhance the model’s learning capacity and ensure steady training of the deep network. The input to our model is a \( d \)-dimensional coordinate vector, which corresponds to an output scalar value. Therefore, our implicit neural network learns a function \( F(\theta): \mathbb{R}^d \rightarrow \mathbb{R}\), where \( \theta \) represents the parameters of the neural network. 

\subsection{Uncertainty Quantification Using MCDropout Method}
The architecture implemented for the MCDropout method is depicted in Fig.~\ref{model_arch}. At the last two residual blocks, a post-activation dropout layer is added to make our model dropout-enabled. This is done to conveniently calculate the prediction uncertainty during inference for the MCDropout method. Additionally, during training, dropout layers also help with regularization. In principle, a dropout layer should be incorporated for each residual block to approximate a fully Bayesian neural network~\cite{segnet}. However, Kendall et al.~\cite{segnet} show that adding dropout at each residual block or after each hidden layer can act as a strong regularizer, potentially reducing the overall prediction accuracy~\cite{segnet}. They further suggest that incorporating dropout to the last layer or a small subset of layers is sufficient to generate high-quality predictions along with reliable uncertainty estimates. Hence, we use dropout layer only at the last two residual blocks to approximate a partial Bayesian neural network to produce accurate predictions as well as robust uncertainty estimates. The impact of using different number of dropout layers on the model performance has been further studied in Section~\ref{evaluation}.

As discussed previously, inference using the MCDropout method involves generating a set of Monte Carlo samples by performing multiple forward passes of the dropout-enabled trained network. Hence,  we generate $m$ instances (realizations) of the scalar field and then calculate the average scalar field, serving as the predicted (expected) scalar field. Typically, the number of samples needed is decided by checking the convergence of the computed expected (averaged) field, meaning that adding more Monte Carlo samples does not improve the reconstruction quality. The uncertainty associated with each grid point is calculated by measuring the standard deviation among the $m$ scalar values. Note that this uncertainty is measured at each grid point in the spatial domain. However, since our focus in this work is to estimate fine-grained uncertainty in the volume rendered images, we further quantify uncertainty in image space using the volume rendering. To estimate the image space pixel-wise uncertainty for enabling uncertainty-aware volume visualization, we compute the uncertainty of the pixel color values after collecting outputs of ray casting algorithm when applied to each individual MC volume realization ($100$ in our experiments). Details of this pixel-wise uncertainty calculation are provided in Section~\ref{pixel_uncert_computation}.

\subsection{Uncertainty Quantification Using the Ensemble Method}
The model architecture for the Ensemble model is the same as shown in Fig.\ref{model_arch}, except no dropout layers are used. Multiple instances of this model are trained to produce a Deep Ensemble model\cite{deepEnsembles}. To create a robust ensemble model, we train \(n\) instances of the SIREN model. The variability across each ensemble member is ensured by randomly shuffling the order of data points at each iteration during training. Once training is completed, the averaged prediction from all \(n\) member models at each grid point produces the expected volume field. Note that the number of ensemble members needed is decided by checking the convergence of the computed expected (averaged) volume, meaning that adding more ensemble member models does not improve the reconstruction quality. Similar to the MCDropout method, we compute the data space uncertainty associated with each grid point. Subsequently, image space pixel-wise uncertainty for the Ensemble method is then computed following the same strategy as the MCDropout method, and is further discussed in Section~\ref{pixel_uncert_computation}.

\begin{figure}[tb]
\centering
\includegraphics[width=\linewidth]{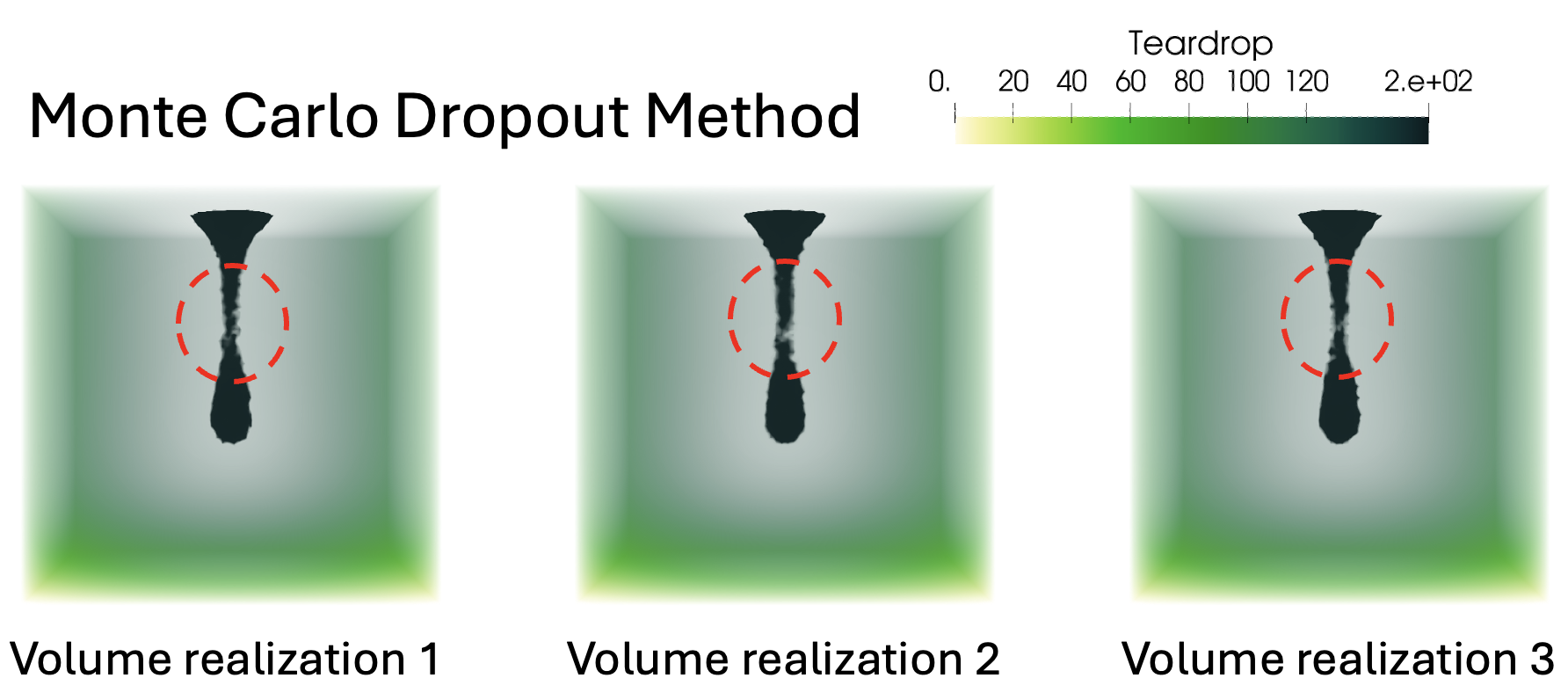}
\includegraphics[width=\linewidth]{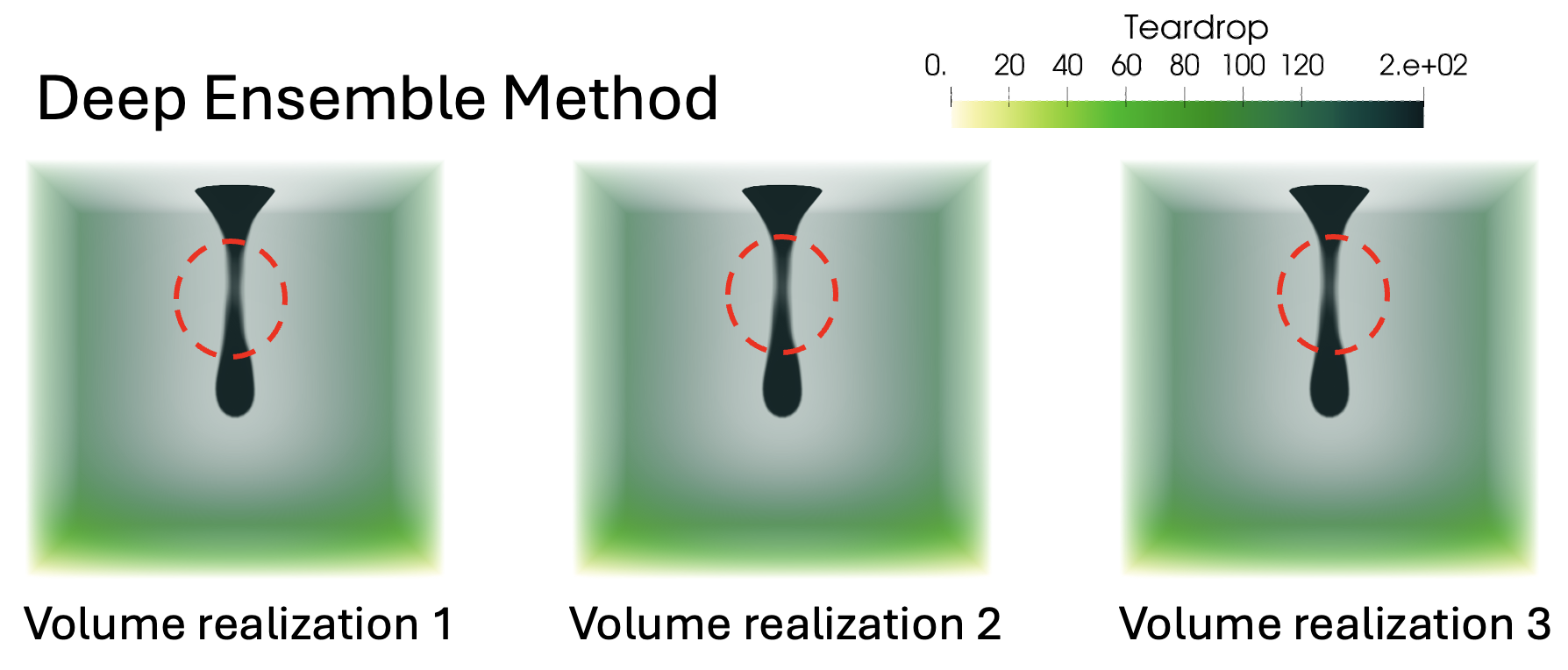}
\caption{Volume visualization of Teardrop data set for three representative MC-sampled fields from the MCDropout method (top row) and three representative fields generated from three ensemble members (bottom row) for the Ensemble method. The ground truth is shown in Fig.~\ref{teardrop_final}. It is observed that individual MC sampled fields produce inaccurate visualization at the thin central segment of the Teardrop data, highlighted by red dotted circles for the MCDropout method. In contrast, the visualization produced by individual ensemble members are more accurate.}
\label{teardrop_samples}
\end{figure}

\begin{figure}[tb]
\centering
\includegraphics[width=\linewidth]{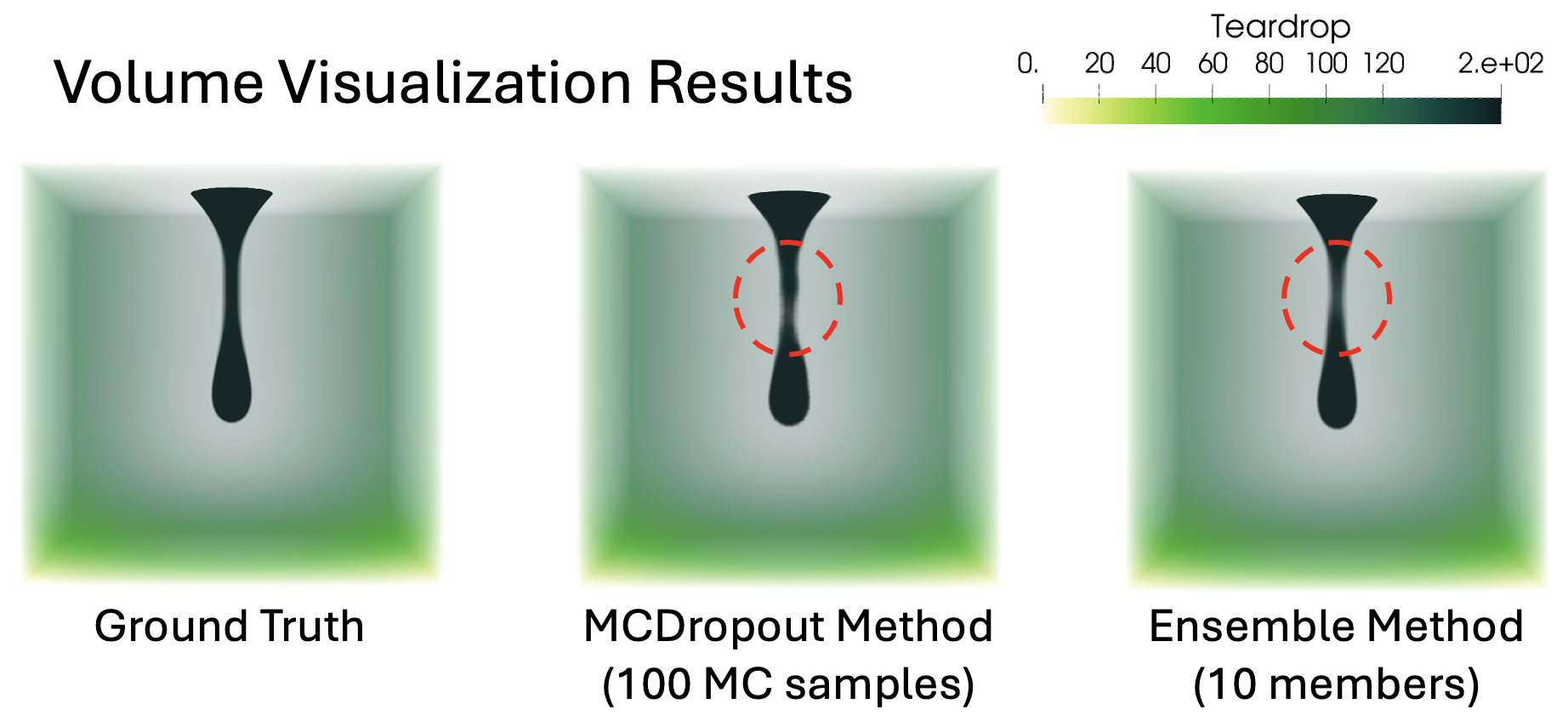}
\caption{Visualization of ground truth, expected (averaged) volume visualization by the MCDropout method, and expected (averaged) volume visualization by the Ensemble method of Teardrop data set. We observe that both MCDropout and Ensemble methods produce comparable and high-quality rendering results.}
\label{teardrop_final}
\end{figure}

\subsection{Loss Function and Hyperparameters}
Our uncertainty-aware SIREN model uses $10$ residual blocks for MCDropout and Ensemble methods. We use $50$ neurons in each hidden layer to generate consistent and comparable results while producing a compact uncertainty-aware representation of the volume data set. Conventional mean squared error loss ($\mathcal{L}_{mse}$) is used to train the network. We use empirical experimentation to identify a suitable learning rate and batch size combination that produces stable, consistent, and high-quality results across all data sets. We utilize a batch size of $2048$ with the Adam optimizer~\cite{kiba14}, setting the learning rate at $0.00005$ and the two Adam optimizer coefficients $\beta_1$ and $\beta_2$ to their default values at $0.9$ and $0.999$, respectively. Furthermore, a learning rate decay mechanism is adopted to optimize the training process, with a decay factor of $0.8$ and a step size of $15$. All the models were trained for $300$ epochs to ensure convergence and robustness in performance evaluation. During training, we use a low dropout rate (probability) of $\eta=0.001$ for Hurricane Isabel and Combustion data sets. This training dropout rate results in a poorly performing MCDropout model for the Teardrop data set. Hence, we use a higher dropout rate of $\eta=0.05$, which produces a stable MCDropout model for the Teardrop data set. During inference, we use a consistent dropout rate of $\eta=0.1$ for all the data sets to generate robust uncertainty estimates. No dropout is used for training ensemble models. This consistent approach in model design ensures a rigorous assessment of the effectiveness of uncertainty-aware deep neural networks across various data sets.  
\begin{table}[thb]
\centering
\caption{Description of data sets that are used in the experimentation.}
\label{datadesc_table}
\resizebox{\linewidth}{!}{
\begin{tabular}{|c|c|c|}
\hline
\textbf{Data set} & \textbf{Dimensionality} & \textbf{Spatial Resolution} \\ \hline
Teardrop & 3D             & 64 $\times$ 64 $\times$ 64             \\ \hline
Isabel Pressure (T=25) & 3D             & 250 $\times$ 250 $\times$ 50              \\ \hline
Isabel Velocity (T=25) & 3D             & 250 $\times$ 250 $\times$ 50  \\ \hline
Combustion Mixfrac (T=41)       & 3D             & 240 $\times$ 360 $\times$ 60\\ \hline
\end{tabular}
}
\end{table}

\section{Uncertainty-aware Volume Visualization}
\label{results}
We perform a comprehensive study of the uncertainty-aware INRs  using four volume data sets. The dimensionality and spatial resolution of these data sets are reported in Table~\ref{datadesc_table}. A GPU server with NVIDIA GeForce GTX $1080$Ti GPUs with $12$GB GPU memory is used for model training and volume reconstruction. The rendering is done on a MacBook Pro with an Apple M1 Pro chip with $10$ CPU and $16$ GPU cores and $16$GB memory. All the models are implemented in PyTorch~\cite{pytorch2019}. Teardrop data set is generated using a mathematical function~\cite{teardropdata} sampled on a $64 \times 64 \times 64$ uniform grid. Hurricane Isabel data was produced by the Weather Research and Forecast model, courtesy of NCAR and the U.S. National Science Foundation (NSF). Turbulent Combustion data set is made available by Dr. Jacqueline Chen at Sandia Laboratory through U.S. Department of Energy’s SciDAC Institute for Ultrascale Visualization.

\begin{figure*}[tb]
\centering
\includegraphics[width=0.75\linewidth]{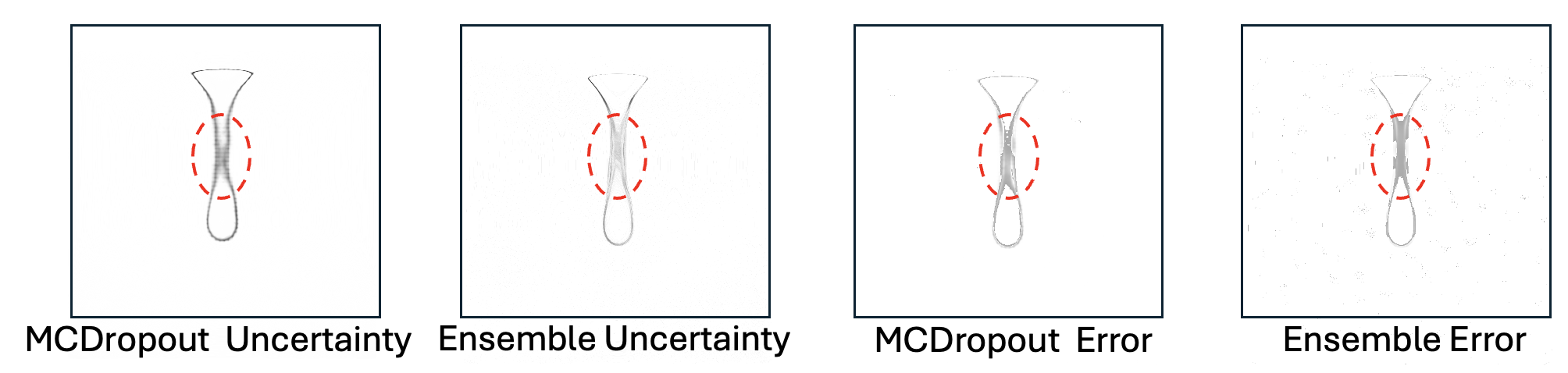}
\caption{Prediction uncertainty and error maps of Teardrop data. Both methods produce high uncertainty and error at the thin central segment (highlighted by red dotted circles). Such uncertainty information can readily help the users to identify regions where the model is under-confident. We further observe that the teardrop's boundary also shows higher uncertainty and error, indicating that both methods are also unable to confidently predict the sharp boundary regions.}
\label{teardrop_total_uncertainty_error}
\end{figure*}

\subsection{Computation of Pixel-wise Prediction Uncertainty for Volume Visualization}
\label{pixel_uncert_computation}
We use the trained INRs to reconstruct the entire volume to comprehensively assess the model’s reconstruction quality, prediction uncertainty, and error. Through empirical experimentation, we observe that $100$ MC samples for the MCDropout method and $10$ ensemble members for the Ensemble method allow us to produce robust estimates of volume data. A study on how reconstruction quality changes given different numbers of MC samples for the MCDropout method and a different number of ensemble members for the Ensemble method is provided later in Section~\ref{evaluation}. Thus, unless specified otherwise, we use $100$ MC samples for MCDropout and $10$ ensemble members for the Ensemble method in all experiments.

\bmark{Our goal is to enable visualization and comprehension of fine-grained prediction uncertainty when model-reconstructed scalar data sets are visualized using volume rendering methods for any user-specified transfer functions. As the rendered image depends on transfer functions, the associated uncertainty map should also be updated when the transfer function changes.} Both uncertainty estimation methods we use require multiple volume realizations (100 for the MCDropout method and 10 for the Ensemble method) to generate the final volume-rendered image and the associated uncertainty map. We utilize individual volume realizations to estimate the fine-grained pixel-wise uncertainty from the model-predicted results. Given a user-specified transfer function and view direction, our method applies the ray casting algorithm to each volume realization (100 for the MCDropout method and 10 for the Ensemble method) and collects the RGB pixel values for each instance. Then, the final result is obtained by averaging the RGB pixel colors, and the associated uncertainty is estimated by computing the pixel-wise standard deviation. We first estimate each color channel's uncertainty (standard deviation) from the volume rendering results. Then, we compute the final mean pixel uncertainty by averaging the individual color channel uncertainty values. The uncertainty values are then normalized and stored as grayscale images, where darker colors indicate higher uncertainty for the corresponding pixel locations. \bmark{As the transfer function changes, this process is repeated so that the updated result and corresponding uncertainty map can be generated for visualization.}

\subsection{Uncertainty-Informed Volume Visualization}
In the following, we qualitatively and visually study the volume visualization results and the estimated uncertainty patterns for MCDropout and Ensemble methods using several volume data sets. Then, in Section~\ref{evaluation}, we provide a quantitative evaluation of volume reconstruction quality and uncertainty estimates for both methods. Next, we further comprehensively evaluate the proposed methods under varying parameter configurations to assess their applicability, usefulness, and implications.

\begin{figure}[tb]
\centering
\includegraphics[width=\linewidth]{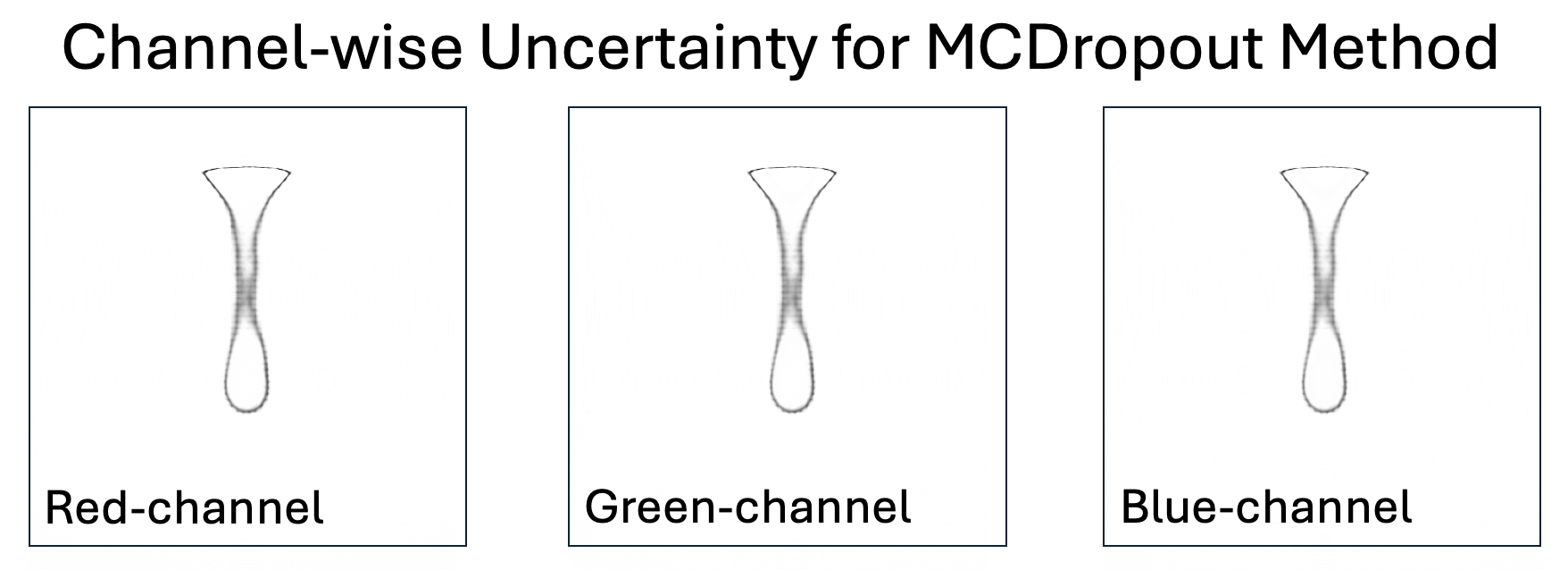}
\includegraphics[width=\linewidth]{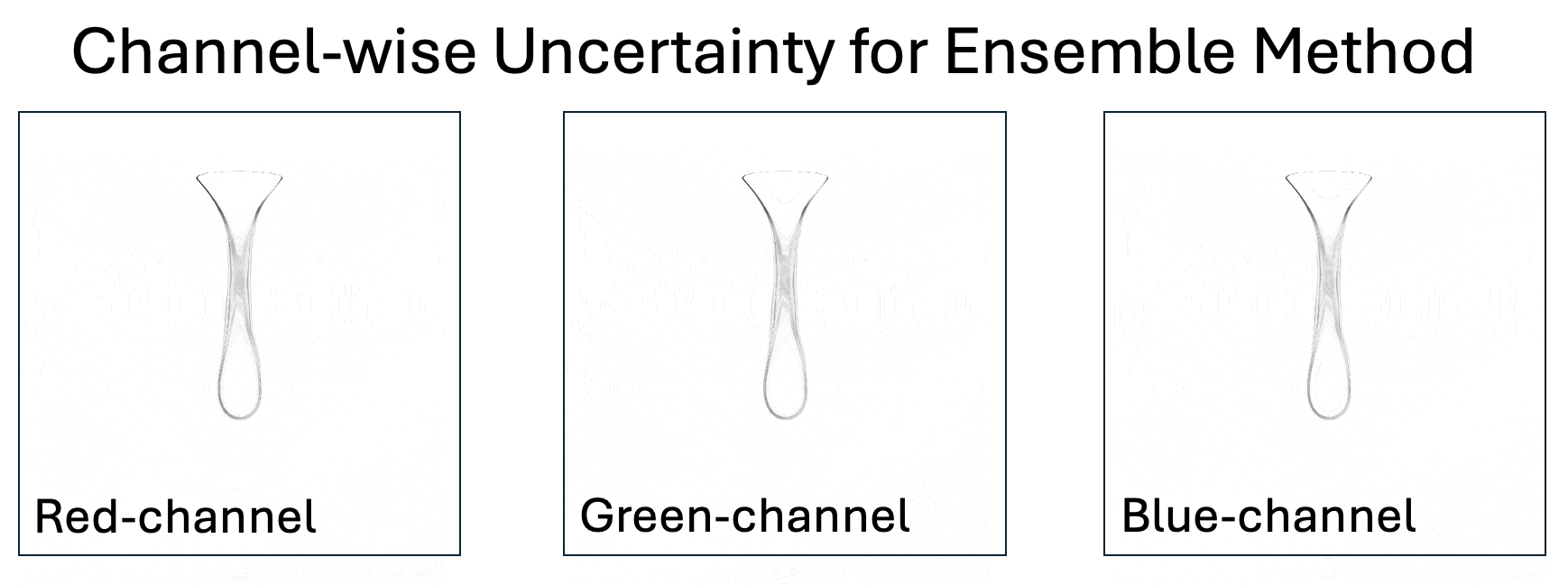}
\caption{Prediction uncertainty visualization of Teardrop data set for individual RGB color channels. We observe that the prediction uncertainty patterns are comparable across all three color channels for the MCDropout and Ensemble methods.}
\label{teardrop_channel_uncert}
\end{figure}

\subsection{Visual Analysis for Teardrop Data}
Our first case study uses the Teardrop data set~\cite{teardropdata}. To study the volume visualization results obtained from the two uncertainty estimation methods, we generate $100$ Monte Carlo sample volume reconstructions for the MCDropout method and $10$ sample volume reconstructions for the ensemble method (as $10$ ensemble members are used). In Fig.~\ref{teardrop_samples}, the top row shows volume rendered results of three individual MC sample volumes for the MCDropout method. Similarly, the bottom row depicts visualization generated by three separate ensemble members. We observe that the individual ensemble members renders the thin central segment region (as highlighted by red dotted circles) more accurately than the individual MCDropout sampled fields when compared against the ground truth rendering shown in Fig.~\ref{teardrop_final}. All the volume visualization results use fixed transfer functions, viewpoints, and all other rendering parameters to ensure fair comparison. 

Next, in Fig.~\ref{teardrop_final}, we show the ground truth rendering, the rendering of the expected (averaged) field constructed using $100$ MC sample fields for the MCDropout method and $10$ ensemble member reconstructed fields for the Ensemble method, respectively. We observe that the expected field generates accurate and visually similar volume rendering results for both methods. Both MCDropout and Ensemble methods accurately preserve the thin central segment, as red dotted circles show.

In Fig.\ref{teardrop_total_uncertainty_error}, we present the pixel-wise prediction uncertainty and error maps for the Teardrop data set using both the MCDropout and Ensemble methods. The uncertainty represents by the pixel-wise standard deviation, averaged over all three color channels. To estimate the pixel-wise standard deviation, we perform ray casting on 100 MC sampled volumes for the MCDropout method and on 10 ensemble member generated volumes for the Ensemble method. The standard deviation for each color channel is computed from the final pixel colors and then averaged to produce the final uncertainty image shown in Fig.\ref{teardrop_total_uncertainty_error}. Darker pixels indicate higher uncertainty. The image is generated by first computing pixel-wise average uncertainty values, then mapping these values to a grayscale colormap where darker colors reflect higher uncertainty. \bmark{The error maps are computed by estimating the channel-wise absolute error between the ground truth pixel colors and the predicted pixel colors. Then the average error is calculated and mapped to a grayscale colormap where darker colors reflect higher error. All uncertainty and error maps are visually comparable as they are generated using a consistent grayscale colormap. Users can compare error and uncertainty between the two uncertainty-estimation methods by comparing the darkness of the pixel intensities.}  It is observed that both methods exhibit higher prediction uncertainty as well as higher error at the thin segment of the teardrop, indicating that the model is less confident and more erroneous in predicting values in this region. Additionally, higher uncertainty and error is noted at the boundary of the teardrop structure, where a sharp change in color gradient is observed, further indicating reduced confidence in boundary predictions. \bmark{The results indicate that the uncertainty and error are largely correlated for this data set.} 

Prediction uncertainty maps for each individual color channel are provided in Fig.~\ref{teardrop_channel_uncert}. The top row displays the results for the MCDropout method, while the bottom row shows the channel-wise uncertainty maps for the Ensemble method. We observe that the estimated uncertainty patterns are identical for each color channel, indicating that each color channel incurs comparable uncertainty estimates in similar spatial regions.   

\subsection{Visual Analysis for Isabel Pressure Field}
\begin{figure*}[tb]
\centering
\includegraphics[width=\linewidth]{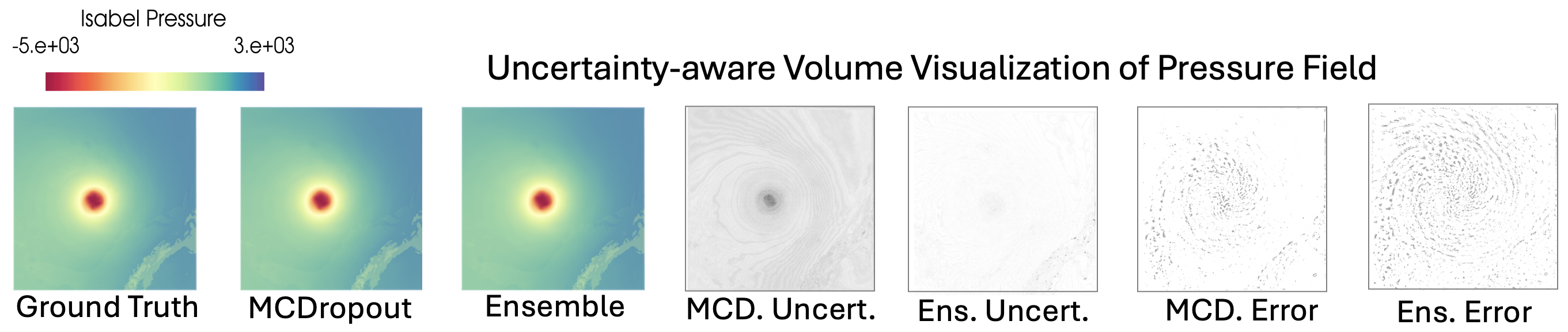}
\caption{Volume visualization of the Pressure field of Hurricane Isabel data set. We observe that both MCDropout and Ensemble methods produce visually identical volume rendering results when compared against the ground truth. The uncertainty maps indicate higher  uncertainty is produced by the MCDropout method at the Hurricane eye region, with moderate uncertainty away from the eye region. However, the Ensemble method produces very confident visualization results with only fewer pixels incurring higher uncertainty near the land region (bottom right corner). The error maps show pixel errors when the predicted image is compared against ground truth. We observe that uncertainty and error maps are not correlated.}
\label{Isabel_pressure_volvis}
\end{figure*}

\begin{figure}[tb]
\centering
\includegraphics[width=\linewidth]{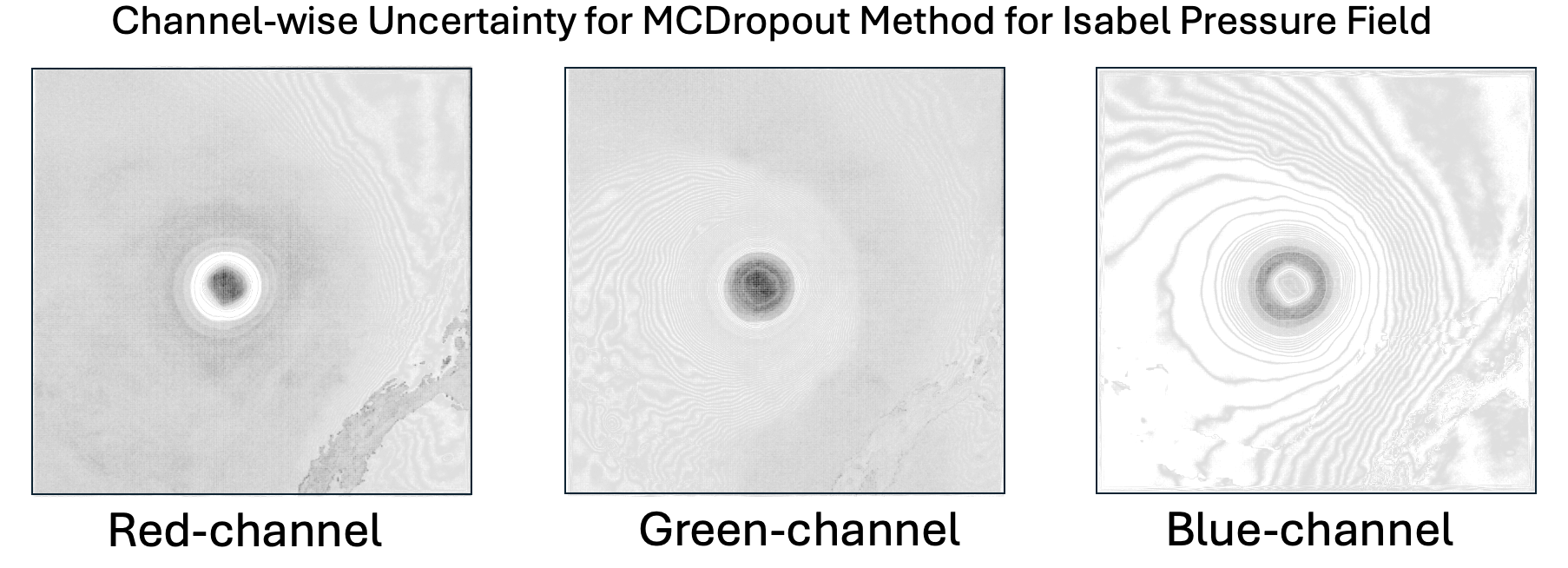}
\includegraphics[width=\linewidth]{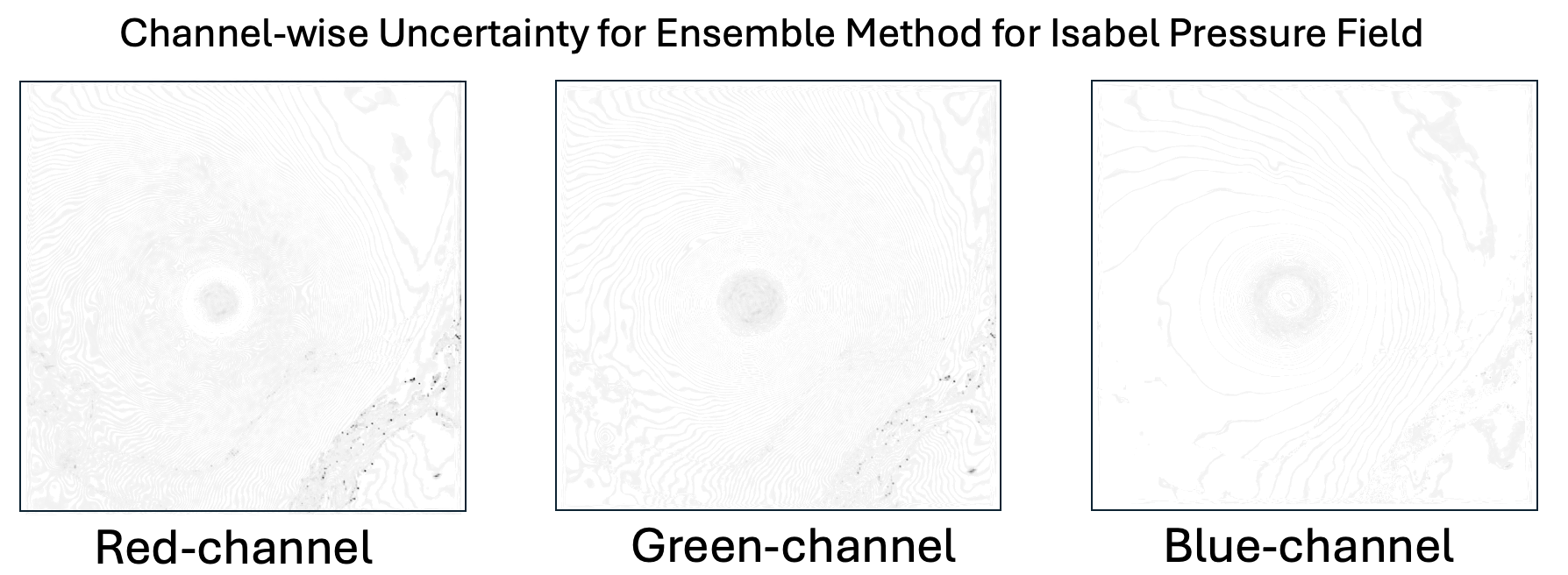}
\caption{Prediction uncertainty visualization of the Pressure field of Hurricane Isabel data set for individual RGB color channels. We observe that the three channels exhibit different uncertainty patterns, with the higher uncertain regions being concentrated around the Hurricane eye region for the MCDropout method. The uncertainty for all three channels for the Ensemble method is comparatively lower, indicating a more robust prediction.}
\label{Isabel_pressure_channel_uncert}
\end{figure}

Our next case study utilizes the Pressure field from the Hurricane Isabel data set. Both the MCDropout and Ensemble methods produce highly accurate visualizations when compared to the ground truth, as shown in Fig.~\ref{Isabel_pressure_volvis}. A close inspection of the uncertainty map generated by the MCDropout method reveals high uncertainty in the Hurricane eye feature region and moderate pixel uncertainty in the surrounding area. In contrast, the Ensemble method produces a cleaner uncertainty map, with higher uncertainty confined to only a few pixels in the land region (bottom right corner). These uncertainty maps indicate that the Ensemble method yields more robust predictions with high confidence, resulting in more reliable volume visualization images compared to the MCDropout method. \bmark{When the two error maps from the two methods are investigated, it is observed that the error and uncertainty maps convey different information about the models and the MCDropout method produces fewer pixels with higher errors.}

In Fig.~\ref{Isabel_pressure_channel_uncert}, we present the channel-wise uncertainty maps for the Isabel Pressure field using both the MCDropout (top row) and Ensemble (bottom row) methods. It is observed that the Hurricane eye region exhibits high prediction uncertainty across all three color channels for the MCDropout method, with moderate uncertainty in the rest of the spatial domain. This indicates that model prediction uncertainty can affect individual color channels differently, and displaying channel-wise uncertainty maps allows for a detailed investigation of the impact of uncertainty on each color channel separately. For the Ensemble method, the overall uncertainty for all three color channels is much lower, with minimal variation in the uncertainty patterns among them.

\subsection{Visual Analysis for Isabel Velocity Magnitude Field}
\begin{figure*}[tb]
\centering
\includegraphics[width=\linewidth]{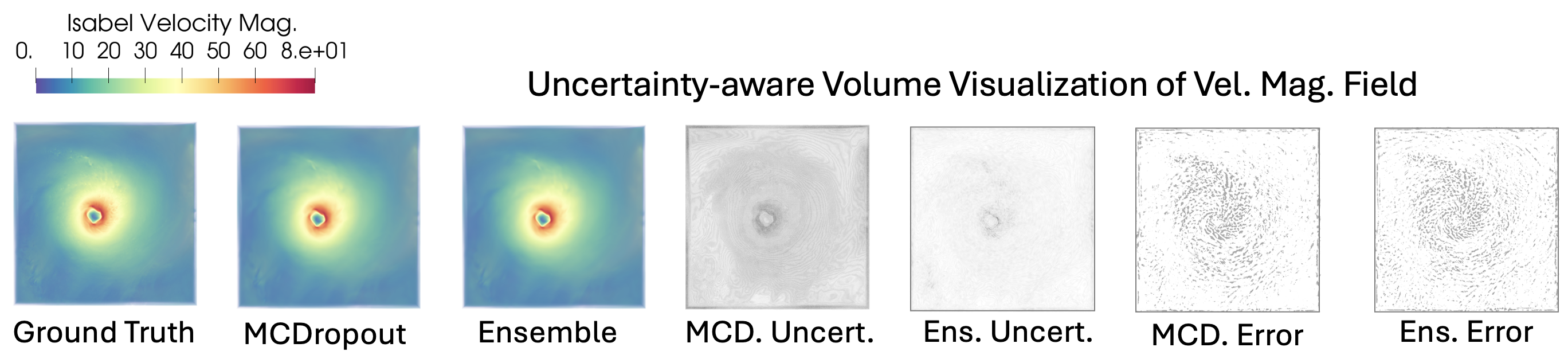}
\caption{Volume visualization of the Velocity Magnitude field of Hurricane Isabel data set. We observe that both MCDropout and Ensemble methods produce visually identical volume rendering results when compared against the ground truth. The uncertainty map indicates moderate uncertainty in the region surrounding the Hurricane eye feature for the MCDropout method. In contrast, the Ensemble method produces an uncertainty map where fewer pixels show high uncertainty compared to the MCDropout method. The error maps show pixel errors when the predicted image is compared against ground truth. We observe similar error maps for both methods while also note that uncertainty and error maps are not correlated.}
\label{isabel_velmag_volvis}
\end{figure*}

Fig.~\ref{isabel_velmag_volvis} illustrates uncertainty-aware volume visualization results for the Velocity Magnitude field of the Hurricane Isabel data set. As with the Pressure field, both the MCDropout and Ensemble methods produce accurate volume visualizations. The uncertainty maps reveal a similar pattern: the MCDropout method shows moderate uncertainty over a broader area around the Hurricane eye feature, while the Ensemble method exhibits higher uncertainty concentrated around the eye wall of the Hurricane. \bmark{The error maps from both methods display similar error patterns, with higher prediction errors concentrated around the Hurricane eye. Examining the uncertainty and error maps of the Ensemble method reveals that, while it predicts the regions around the Hurricane eye with greater confidence, it can still make errors in these areas. This indicates instances where the model is making overconfident predictions.}

\subsection{Visual Analysis for Combustion Mixfrac Field}
\begin{figure*}[tb]
\centering
\includegraphics[width=\linewidth]{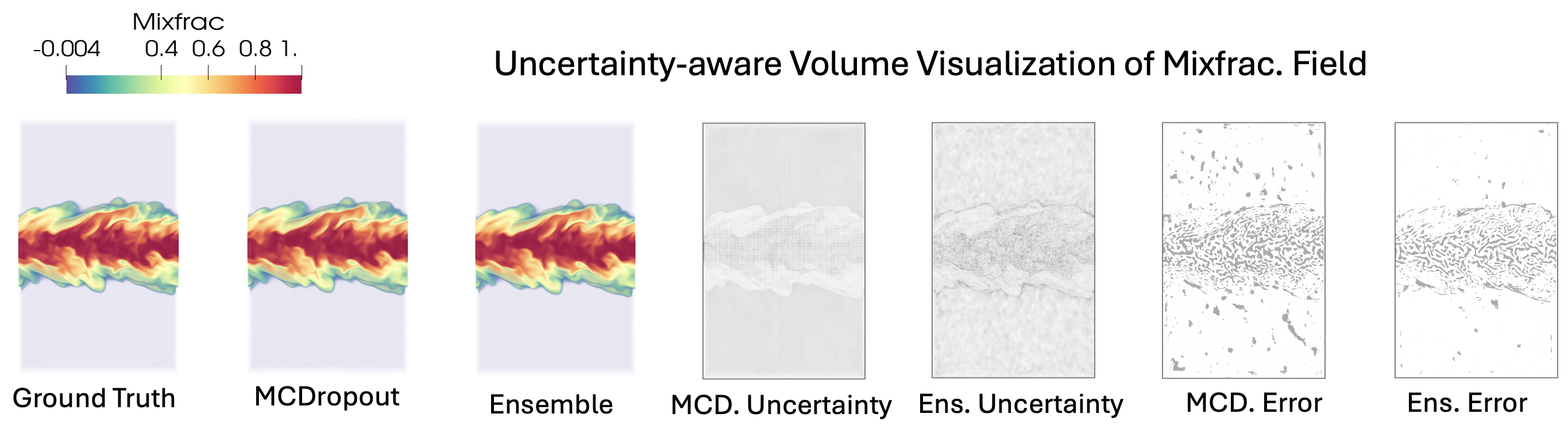}
\caption{Uncertainty-informed volume visualization of the Mixfrac field of Turbulent Combustion data set. We observe that both MCDropout and Ensemble methods produce visually identical volume rendering results when compared against the ground truth. The uncertainty maps show that while the MCDropout method produces smoother and moderate to low uncertainty, the Ensemble method results in noisier and higher uncertainty inside and around the burning flame region. The two error maps indicate that the MCDropout method incurs more error in pixel values as compared to the Ensemble method.}
\label{Comb_mixfrac_volvis}
\end{figure*}

Finally, we present the visualization results for the Mixture Fraction (Mixfrac) field of the Turbulent Combustion data set in Fig.~\ref{Comb_mixfrac_volvis}. Both methods produce visually identical volume rendering results for the Mixfrac field when compared to the ground truth. However, a comparison of the two uncertainty maps, generated by the MCDropout and Ensemble methods, reveals that the MCDropout method produces a smoother uncertainty map, while the Ensemble method yields a relatively noisier uncertainty map. Additionally, the Ensemble method shows high uncertainty in the turbulent flame regions, whereas the MCDropout method results in lower prediction uncertainty, indicating more robust predictions for such regions. \bmark{A close examination of the two error maps reveals that both methods produce similar patterns, with higher prediction errors predominantly occurring in the complex burning regions.}

\section{Evaluation and Parameter Study}
\label{evaluation}

\subsection{Reconstruction Quality and Prediction Error in Volume Space}
Table~\ref{recon_table} presents a comparative study between the MCDropout and Ensemble methods in terms of averaged (expected) volume reconstruction quality measured using Peak signal-to-noise-ratio (PSNR). We use predictions from $100$ MC samples for MCDropout and $10$ ensemble members for Ensemble method to compute the final averaged field. Besides PSNR, we also estimate the Root Mean Squared Error (RMSE) between the reconstructed and ground truth fields. We observe that the reconstruction quality and RMSE between the MCDropout method and a single model trained without any dropout layer are comparable. However, the Ensemble method consistently produces the best reconstruction quality (highest PSNR) with minimum RMSE for all the data sets.
\begin{table}[thb]
\centering
\caption{PSNR (dB) ($\uparrow$) and RMSE ($\downarrow$) values for different methods (No Dropout, MCDropout with 100 samples, and Ensemble with 10 members) applied over data sets. We observe  Ensemble method gets higher PSNR and lower RMSE consistently, indicating better performance in reconstruction.}
\label{recon_table}
\resizebox{\linewidth}{!}{
\begin{tabular}{|c|c|c|c|}
\hline
\textbf{Data set} & \textbf{Method} & \textbf{PSNR (dB)} & \textbf{RMSE} \\ \hline
\multirow{3}{*}{\textbf{Teardrop}} & No Dropout & 72.009 & 0.040 \\ \cline{2-4} 
 & MCDropout (100 MC samples) & 71.694 & 0.042 \\ \cline{2-4} 
 & Ensemble (10 members) & 76.832 & 0.023 \\ \hline
\multirow{3}{*}{\textbf{Isabel Pressure}} & No Dropout & 56.935 & 10.683 \\ \cline{2-4} 
 & MCDropout (100 MC samples) & 56.922 & 11.116 \\ \cline{2-4} 
 & Ensemble (10 members) & 57.068 & 10.542 \\ \hline
\multirow{3}{*}{\textbf{Isabel Velocity}} & No Dropout & 44.324 & 0.469 \\ \cline{2-4} 
 & MCDropout (100 MC samples) & 44.178 & 0.475 \\ \cline{2-4} 
 & Ensemble (10 members) & 44.717 & 0.451 \\ \hline
\multirow{3}{*}{\textbf{Combustion Mixfrac}} & No Dropout & 48.779 & 0.004 \\ \cline{2-4} 
 & MCDropout (100 MC samples) & 48.475 & 0.004 \\ \cline{2-4} 
 & Ensemble (10 members) & 49.434 & 0.003 \\ \hline
\end{tabular}
}
\end{table}

\subsection{PSNR Study with Varying Number of Ensemble Members} Table~\ref{ensemble_samples_psnr} illustrates the effect of varying the number of ensemble members on PSNR. Notably, achieving robust predictions requires fewer ensemble members compared to the MCDropout method. The data shows that increasing the number of ensemble members slowly improves the PSNR and essentially leading the PSNR value to become saturated for $10$ ensemble members. Nevertheless, for consistency and robustness  across our experiments, we utilize $10$ ensemble members.
\begin{table}[thb]
\centering
\caption{PSNR (dB) ($\uparrow$) when the varying number of Ensemble members are used for computation in the Ensemble method.}
\label{ensemble_samples_psnr}
\resizebox{\linewidth}{!}{
\begin{tabular}{|c|c|c|c|c|}
\hline
\textbf{Data set} & \textbf{\begin{tabular}[c]{@{}c@{}}\#Ens mem \\ = 2\end{tabular}} & \textbf{\begin{tabular}[c]{@{}c@{}}\#Ens mem \\ = 5\end{tabular}} & \textbf{\begin{tabular}[c]{@{}c@{}}\#Ens mem \\ = 7\end{tabular}} & \textbf{\begin{tabular}[c]{@{}c@{}}\#Ens mem \\ = 10\end{tabular}} \\ \hline
\textbf{Teardrop} & 76.321 & 76.682 & 76.738 & 76.778 \\ \hline
\textbf{Isabel Pressure} & 57.838 & 58.176 & 58.321 & 58.425 \\ \hline
\textbf{Isabel Velocity} & 44.812 & 45.174 & 45.275 & 45.372 \\ \hline
\textbf{Combustion Mixfrac} & 49.931 & 51.723 & 51.997 & 52.401 \\ \hline
\end{tabular}
}
\end{table}

\subsection{PSNR Study with Varying Number of MC Samples}
Table~\ref{MC_samples_psnr} shows the PSNR values of the scalar obtained by averaging different numbers of MC samples. Increasing the number of MC samples up to 100 results in a slight PSNR improvement. Beyond this point, the PSNR gains become marginal, making it impractical to consider more samples. Therefore, to balance computation time and prediction quality, we use $100$ MC samples for all experiments involving the MCDropout method.
\begin{table}[thb]
\centering
\caption{PSNR (dB) ($\uparrow$) when the varying number of MC samples are used for computation in the MCDropout method.}
\label{MC_samples_psnr}
\resizebox{\linewidth}{!}{
\begin{tabular}{|c|c|c|c|c|c|}
\hline
\textbf{Data set} & \textbf{\begin{tabular}[c]{@{}c@{}}\#MCSamp\\ =10\end{tabular}} & \textbf{\begin{tabular}[c]{@{}c@{}}\#MCSamp\\ =25\end{tabular}} & \textbf{\begin{tabular}[c]{@{}c@{}}\#MCSamp\\ =50\end{tabular}} & \textbf{\begin{tabular}[c]{@{}c@{}}\#MCSamp\\ =75\end{tabular}} & \textbf{\begin{tabular}[c]{@{}c@{}}\#MCSamp\\ =100\end{tabular}} \\ \hline
\textbf{Teardrop} & 69.264 & 70.749 & 71.425 & 71.663 & 71.782 \\ \hline
\textbf{Isabel Pressure} & 54.679 & 55.869 & 56.356 & 56.530 & 56.922 \\ \hline
\textbf{Isabel Velocity} & 43.671 & 43.975 & 44.081 & 44.117 & 44.135 \\ \hline
\textbf{Combustion Mixfrac} & 47.165 & 47.967 & 48.270 & 48.375 & 48.428 \\ \hline
\end{tabular}
}
\end{table}

\subsection{Impact of Different Number of Ensemble Members on Average Pixel-wise Image space Uncertainty} Table~\ref{pixel_uncert_ens} illustrates the impact of varying the number of ensemble members on the average pixel-wise uncertainty value computed in the image space using the volume rendered pixel color values. This value reflects the normalized estimated uncertainty (standard deviation) calculated from the RGB color channels using  the Ensemble method. We observe that the uncertainty value gradually saturates as the number of ensemble members increases up to $10$.
\begin{table}[thb]
\centering
\caption{Image space uncertainty values with different number of ensemble members. We observe that by increasing the number of ensemble members, the uncertainty value tends to become saturated.}
\label{pixel_uncert_ens}
\resizebox{\linewidth}{!}{
\begin{tabular}{|c|c|c|c|c|}
\hline
\textbf{Data set} & \textbf{\begin{tabular}[c]{@{}c@{}}\#Ens mem \\ = 2\end{tabular}} & \textbf{\begin{tabular}[c]{@{}c@{}}\#Ens mem \\ = 5\end{tabular}} & \textbf{\begin{tabular}[c]{@{}c@{}}\#Ens mem \\ = 7\end{tabular}} & \textbf{\begin{tabular}[c]{@{}c@{}}\#Ens mem \\ = 10\end{tabular}} \\ \hline
\textbf{Teardrop} & 0.018 & 0.014 & 0.012 & 0.010 \\ \hline
\textbf{Isabel Pressure} & 0.038 & 0.067 & 0.073 & 0.078 \\ \hline
\textbf{Isabel Velocity} & 0.091 & 0.155 & 0.169 & 0.179 \\ \hline
\textbf{Combustion Mixfrac} & 0.209 & 0.329 & 0.351 & 0.365 \\ \hline
\end{tabular}
}
\end{table}

\subsection{Impact of Different Number of Dropout Layers on MCDropout Model Performance} 
We conduct experiments on the Isabel Pressure and Teardrop data sets to assess the detailed impact of varying the number of dropout layers on model performance and prediction accuracy. We present results of averaged volume reconstruction quality results using models with dropout added at the (1) last two residual blocks, (2) last half of the blocks, and (3) all the residual blocks. The results, shown in Table~\ref{drop_layer_exp}, show that with an increased number of dropout layers, the PSNR value drops slowly, as was reported by Kendall et al.~\cite{segnet} since an increased number of dropout layers can act as a strong regularizer. Hence, in our work, we use dropout at the last two residual blocks to produce high-quality volume reconstruction and robust uncertainty estimates.
\begin{table}[thb]
\centering
\caption{We report how the reconstruction quality (PSNR ($\uparrow$)) changes when different numbers of Dropout layers are used for the MCDropout method. Three configurations are studied: Dropout used (1) at the last two residual blocks; (2) at the last half of the residual blocks; (3) at all the residual blocks.}
\label{drop_layer_exp}
\resizebox{0.8\linewidth}{!}{
\begin{tabular}{|c|c|c|cl|}
\hline
\textbf{Data Set} &
  \begin{tabular}[c]{@{}c@{}}\textbf{Last two} \\ \textbf{Res. Block}\end{tabular} &
  \begin{tabular}[c]{@{}c@{}}\textbf{Last half} \\ \textbf{Res. Block}\end{tabular} &
  \multicolumn{2}{c|}{\begin{tabular}[c]{@{}c@{}}\textbf{All} \\ \textbf{Res. Block}\end{tabular}} \\ \hline
Teardrop &
  71.694 &
  68.048 &
  \multicolumn{2}{c|}{62.158} \\ \hline
Isabel Pressure &
  56.922 &
  56.784 &
  \multicolumn{2}{c|}{56.189} \\ \hline
\end{tabular}
}
\end{table}

\subsection{Reconstruction with Different Dropout Probabilities} 
Table~\ref{varied_dropout_prob} presents a study assessing the quality of volume data reconstruction, measured by PSNR, averaged over $100$ MC samples across different test time dropout probabilities. We observe that PSNR mostly remains stable up to the dropout probability of $0.2$, and as the dropout probability is further increased, expectedly, the PSNR value gradually drops. In their work, Gal et al.~\cite{gagh16} suggest that using a small test time dropout probability is sufficient to estimate the model uncertainty robustly. Therefore, in our experiments, we use a dropout probability of $0.1$ consistently for all the data sets to generate robust and meaningful uncertainty estimates.

\begin{table}[thb]
\centering
\caption{PSNR (dB) ($\uparrow$) of $100$ MC samples with different test time dropout probabilities during inference.}
\label{varied_dropout_prob}
\resizebox{\linewidth}{!}{
\begin{tabular}{|c|cccccc|}
\hline
\multirow{2}{*}{\textbf{Data Set}} &
  \multicolumn{6}{c|}{\textbf{Different Test Time Dropout Probability}} \\ \cline{2-7} 
 &
  \multicolumn{1}{c|}{\textbf{0.05}} &
  \multicolumn{1}{c|}{\textbf{0.1}} &
  \multicolumn{1}{c|}{\textbf{0.2}} &
  \multicolumn{1}{c|}{\textbf{0.3}} &
  \multicolumn{1}{c|}{\textbf{0.4}} &
  \textbf{0.5} \\ \hline
\textbf{Teardrop} &
  \multicolumn{1}{c|}{71.878} &
  \multicolumn{1}{c|}{71.694} &
  \multicolumn{1}{c|}{71.282} &
  \multicolumn{1}{c|}{70.567} &
  \multicolumn{1}{c|}{69.649} &
  68.437 \\ \hline
\textbf{\begin{tabular}[c]{@{}c@{}}Isabel \\ Pressure\end{tabular}} &
  \multicolumn{1}{c|}{56.789} &
  \multicolumn{1}{c|}{56.922} &
  \multicolumn{1}{c|}{56.084} &
  \multicolumn{1}{c|}{55.295} &
  \multicolumn{1}{c|}{54.23} &
  52.728 \\ \hline
\textbf{\begin{tabular}[c]{@{}c@{}}Isabel \\ Velocity\end{tabular}} &
  \multicolumn{1}{c|}{44.268} &
  \multicolumn{1}{c|}{44.178} &
  \multicolumn{1}{c|}{43.924} &
  \multicolumn{1}{c|}{43.45} &
  \multicolumn{1}{c|}{42.788} &
  41.843 \\ \hline
\textbf{\begin{tabular}[c]{@{}c@{}}Combustion\\  Mixfrac\end{tabular}} &
  \multicolumn{1}{c|}{48.658} &
  \multicolumn{1}{c|}{48.474} &
  \multicolumn{1}{c|}{47.855} &
  \multicolumn{1}{c|}{46.848} &
  \multicolumn{1}{c|}{45.45} &
  43.674 \\ \hline
\end{tabular}
}
\end{table}

\subsection{Impact of Different Number of MC Samples on Average Pixel-wise Image space Uncertainty}  
Table~\ref{pixel_uncert_mc} presents the impact of varying the number of MC samples on the average pixel-wise uncertainty value computed in the image space using the volume rendered pixel color values. This value indicates the normalized estimated uncertainty (standard deviation) calculated from the RGB color channels using  the MCDropout method. It is seen that the uncertainty value gradually saturates as the number of MC samples reach $100$.
\begin{table}[thb]
\centering
\caption{Image space uncertainty values with different number of MC samples. We observe that by increasing the number of MC samples up to $100$, the uncertainty values tend to become saturated.}
\label{pixel_uncert_mc}
\resizebox{\linewidth}{!}{
\begin{tabular}{|c|c|c|c|c|c|}
\hline
\textbf{Data Set} &
  \textbf{\begin{tabular}[c]{@{}c@{}}\#MCSamp\\ =10\end{tabular}} &
  \textbf{\begin{tabular}[c]{@{}c@{}}\#MCSamp\\ =25\end{tabular}} &
  \textbf{\begin{tabular}[c]{@{}c@{}}\#MCSamp\\ =50\end{tabular}} &
  \textbf{\begin{tabular}[c]{@{}c@{}}\#MCSamp\\ =75\end{tabular}} &
  \textbf{\begin{tabular}[c]{@{}c@{}}\#MCSamp\\ =100\end{tabular}} \\ \hline
\textbf{Teardrop}                                                      & 0.213 & 0.231 & 0.239 & 0.239                         & 0.24  \\ \hline
\textbf{\begin{tabular}[c]{@{}c@{}}Isabel \\ Pressure\end{tabular}}    & 0.199 & 0.213 & 0.217 & 0.219 & 0.22  \\ \hline
\textbf{\begin{tabular}[c]{@{}c@{}}Isabel \\ Velocity\end{tabular}}    & 0.245 & 0.261 & 0.266 & 0.268                         & 0.269 \\ \hline
\textbf{\begin{tabular}[c]{@{}c@{}}Combustion \\ Mixfrac\end{tabular}} & 0.637 & 0.671 & 0.681 & 0.685                         & 0.686 \\ \hline
\end{tabular}
}
\end{table}

\subsection{Reconstruction Quality and Error in Image Space} Table~\ref{image_space_psnr_error} provides the PSNR and RMSE values computed using the averaged volume-rendered images for both the methods. The PSNR value reflects the reconstruction quality, while the RMSE indicates the error when the ground truth image pixel values are compared against the averaged image for both methods. The detailed computation of these averaged volume-rendered images has been discussed in Section~\ref{pixel_uncert_computation}. A close inspection of Table~\ref{image_space_psnr_error} reveals that the Ensemble method generally produces higher quality volume-rendered images with lower RMSE values.
\begin{table}[thb]
\centering
\caption{Image space PSNR(dB) ($\uparrow$) and RMSE ($\downarrow$) values for MCDropout  (100 samples) and Ensemble method (10 members). We observe that the Ensemble method generally produce higher PSNR and lower RMSE values.}
\label{image_space_psnr_error}
\resizebox{\linewidth}{!}{
\begin{tabular}{|c|c|c|c|}
\hline
\textbf{Data set} & \textbf{Method} & \textbf{PSNR (dB)} & \textbf{RMSE} \\ \hline
\multirow{3}{*}{\textbf{Teardrop}} & \multirow{2}{*}{MC-Dropout (100 MC samples)} & \multirow{2}{*}{48.68} & \multirow{2}{*}{0.882} \\
 &  &  &  \\ \cline{2-4} 
 & Ensemble (10 members) & 49.238 & 0.822 \\ \hline
\multirow{3}{*}{\textbf{Isabel Pressure}} & \multirow{2}{*}{MC-Dropout (100 MC samples)} & \multirow{2}{*}{53.344} & \multirow{2}{*}{0.384} \\
 &  &  &  \\ \cline{2-4} 
 & Ensemble (10 members) & 56.050 & 0.525 \\ \hline
\multirow{3}{*}{\textbf{Isabel Velocity}} & \multirow{2}{*}{MC-Dropout (100 MC samples)} & \multirow{2}{*}{48.122} & \multirow{2}{*}{0.824} \\
 &  &  &  \\ \cline{2-4} 
 & Ensemble (10 members) & 49.072 & 0.739 \\ \hline
\multirow{3}{*}{\textbf{Combustion Mixfrac}} & \multirow{2}{*}{MC-Dropout (100 MC samples)} & \multirow{2}{*}{48.394} & \multirow{2}{*}{0.959} \\
 &  &  &  \\ \cline{2-4} 
 & Ensemble (10 members) & 51.431 & 0.676 \\ \hline
\end{tabular}
}
\end{table}

\begin{table}[thb]
\centering
\caption{ Training and inference timings for MCDropout and Ensemble Methods. During reconstruction, we use 100 MC samples for the MCDropout method and 10 ensemble member outputs for the Ensemble method.}
\label{timing_table}
\resizebox{\linewidth}{!}{
\begin{tabular}{|c|cc|cccc|}
\hline
\multirow{3}{*}{\textbf{Data Set}} & \multicolumn{2}{c|}{\textbf{Training Time (Hours)}} & \multicolumn{4}{c|}{\textbf{Inference time (Seconds)}} \\ \cline{2-7} 
 & \multicolumn{2}{l|}{} & \multicolumn{2}{c|}{\textbf{Reconstruction time}} & \multicolumn{2}{c|}{\textbf{Rendering time}} \\ \cline{2-7} 
 & \multicolumn{1}{c|}{\textbf{MCD}} & \textbf{\begin{tabular}[c]{@{}c@{}}ENS\end{tabular}} & \multicolumn{1}{c|}{\textbf{\begin{tabular}[c]{@{}c@{}}MCD\end{tabular}}} & \multicolumn{1}{c|}{\textbf{\begin{tabular}[c]{@{}c@{}}ENS\end{tabular}}} & \multicolumn{1}{c|}{\textbf{\begin{tabular}[c]{@{}c@{}}MCD\end{tabular}}} & \textbf{\begin{tabular}[c]{@{}c@{}}ENS\end{tabular}} \\ \hline
\textbf{Teardrop} & \multicolumn{1}{c|}{0.42} & 0.43 & \multicolumn{1}{c|}{28.95} & \multicolumn{1}{c|}{2.58} & \multicolumn{1}{c|}{49.902} & 0.838 \\ \hline
\textbf{\begin{tabular}[c]{@{}c@{}}Isabel \\ Pressure\end{tabular}} & \multicolumn{1}{c|}{5.24} & 5.36 & \multicolumn{1}{c|}{328.14} & \multicolumn{1}{c|}{31.86} & \multicolumn{1}{c|}{24.586} & 0.779 \\ \hline
\textbf{\begin{tabular}[c]{@{}c@{}}Isabel \\ Velocity\end{tabular}} & \multicolumn{1}{c|}{5.18} & 5.18 & \multicolumn{1}{c|}{330.91} & \multicolumn{1}{c|}{31.64} & \multicolumn{1}{c|}{38.554} & 0.945 \\ \hline
\textbf{\begin{tabular}[c]{@{}c@{}}Combustion \\ Mixfrac\end{tabular}} & \multicolumn{1}{c|}{8.71} & 8.54 & \multicolumn{1}{c|}{545.56} & \multicolumn{1}{c|}{31.63} & \multicolumn{1}{c|}{40.403} & 1.183 \\ \hline
\end{tabular}
}
\end{table}

\subsection{Comparison of Training and Inference Time Between MCDropout and Ensemble Methods} Table~\ref{timing_table} presents the training and inference times for both methods. For inference, we report both the volume reconstruction time and the ray casting time for each method. The Ensemble approach has faster inference times since it uses predictions from only $10$ members, whereas MCDropout generates 100 MC samples, resulting in longer inference times. Training a dropout-enabled MCDropout model and a single ensemble member takes a comparable amount of time. However, since we train 10 ensemble members to build a robust Ensemble model, the total training time is $10$ times greater, making the Ensemble method computationally more expensive than the MCDropout method. Additionally, storing model checkpoints for the Ensemble method requires $10$ times more storage than for the MCDropout method. This significant difference in training time often makes MCDropout preferable for achieving timely results with robust uncertainty estimates.

\subsection{Model vs. Raw Data Size}
\begin{table}[thb]
\centering
\caption{Comparison of model size and raw data size.}
\label{model_data_size}
\resizebox{\linewidth}{!}{
\begin{tabular}{|c|c|c|c|}
\hline
\textbf{Data Set} & \textbf{Raw Data Size (KB)} & \textbf{\begin{tabular}[c]{@{}c@{}}MCDropout Model\\  Size (KB)\end{tabular}} & \textbf{\begin{tabular}[c]{@{}c@{}}Ensemble Model Size\\  (10 members) (KB)\end{tabular}} \\ \hline
\textbf{Teardrop} & 2048 & 216 & 2160 \\ \hline
\textbf{Isabel Pressure} & 12208 & 220 & 2200 \\ \hline
\textbf{Isabel Velocity} & 12208 & 220 & 2200 \\ \hline
\textbf{Combustion Mixfrac} & 20252 & 220 & 2200 \\ \hline
\end{tabular}
}
\end{table}
\bmark{As Implicit Neural Representations (INRs) like ours have proven effective for compactly representing large-scale volumetric data~\cite{levine_neural_compression}, Table~\ref{model_data_size} compares the sizes of the models and the raw data. As expected, MCDropout requires approximately ten times less storage than the Ensemble method, which consists of ten individual models, whereas MCDropout is a single model-based approach. Thus, when both data compression and uncertainty estimation are required, the MCDropout method offers a significantly higher compression ratio.}

\section{Discussion}
\label{discussion}

\textbf{Uncertainty-agnostic vs. Uncertainty-aware DNNs.} In this work, we advocate the use of deep neural networks (DNNs) that can quantify their prediction uncertainty when employed to perform scientific volumetric data visualization tasks. As demonstrated here, these uncertainty-aware neural networks provide important insights into the reliability of their predictions. By effectively communicating this uncertainty to experts, they can make more informed decisions regarding the data features from the visualization results. Integrating uncertainty is also essential for fostering trust in DNN-predicted results used in scientific visualization research.

\textbf{MCDropout vs. Ensemble Method.} Our research explores two approaches for estimating uncertainty using Implicit Neural Representations (INRs) as the base DNN architecture to visualize volume data sets. Despite its long training times, we employ Deep Ensemble, a well-recognized standard, and investigate the single-model-based MCDropout method to address computational challenges. MCDropout is selected for its theoretical elegance and ease of integration into existing DNN models with minimal modifications. Our results indicate that both methods can offer meaningful prediction uncertainty, while the Ensemble method generally produces more accurate and robust predictions with higher confidence compared to the MCDropout method. It is also observed that the uncertainty maps generated by these two methods could be similar (for the Teardrop data set) as well as different and vary across different data sets. However, considering the extremely high training time required to build a robust ensemble model, MCDropout can be a reliable alternative to the Ensemble method when computational resources are constrained, and prompt uncertainty estimation is essential.

\textbf{Error vs. Uncertainty.} Implicit neural models like ours have been proven to be highly effective in generating compressed representations of large-scale volume data sets~\cite{levine_neural_compression}. It has been demonstrated that efficient volume visualization can be achieved using only the compressed INR model-reconstructed data, eliminating the need to access the raw volume data. In these applications, quantifying error becomes challenging when the entire visual analysis relies on model-generated data, as the raw data may not be available. Our uncertainty-aware INRs can enhance the trustworthiness of these model-generated results by offering prediction uncertainty and model confidence information to domain experts.
 
\section{Conclusions and Future Work}
\label{conclusion}
This paper underscores the significance of understanding uncertainty by applying two deep uncertainty estimation methods. We investigate how prediction uncertainty can benefit DNN-based volume data modeling and visualization tasks. Looking ahead, our research aims to extend to time-varying and multivariate volumetric data sets. Additionally, we plan to explore alternative single-model-based deep uncertainty estimation techniques that are computationally efficient for interactive uncertainty-aware volume visualization. The insights gained from uncertainty estimates can pinpoint areas that require targeted training and highlight model limitations in specific data regions. In critical scenarios, the confidence a model has in its predictions is vital, as acknowledging uncertainties fosters greater trust in the model.


\bibliographystyle{abbrv-doi}

\bibliography{template}
\end{document}